Short Paper*

# Designing and Implementing e-School Systems: An Information Systems Approach to School Management of a Community College in Northern Mindanao, Philippines

Benzar Glen S. Grepon
Program Head, College of Computer Studies, Northern Bukidnon State College, Philippines
ben.it2c@gmail.com
(corresponding author)

Niño T. Baran
College of Computer Studies, Northern Bukidnon State College, Philippines

Kenn Migan Vincent C. Gumonan
College of Computer Studies, Northern Bukidnon State College, Philippines

Aldwin Lester M. Martinez
College of Computer Studies, Northern Bukidnon State College, Philippines

Mona Liel E. Lacsa
College of Computer Studies, Northern Bukidnon State College, Philippines

**Abstract**

*Purpose* – Colleges and Universities have been established to provide educational services to the people. Like any other organization, the school has processes and procedures similar to business or industry that involve admissions, processing of data, and generation of reports. Those processes are made possible through a centralized system in storing, processing, and retrieval of data and information, the majority of the schools in the country are already adopting computer-based systems to address their needs especially on their student and school-related transactions. The absence of a computer system and the complexity of the transactions of the college which makes the personnel be loaded with paper works in storing and keeping student records and information is the motivating factor why the School Management Information System has been designed and developed for a community college in the northern part of Mindanao.

*Method* - This paper discusses the Major Functionalities and Modules of the system through its implementation methodology which is the Agile Model and its impact on the delivery of services and procedures in the overall operation of the college.

*Results* – The project has been evaluated based on ISO 25010, a quality model used for product/software quality evaluation systems. Based on the results of the evaluation, SMIS has been Functional, Usable, and Reliable with an average for every criterion above 4.04 indicating very good performance based on a Likert scale descriptive interpretation.

*Conclusion* – Based on the preceding findings of the study, the respondents agreed that the developed e-school system was functional and lifted the transaction process of the school. The faculty and staff have benefited from making use of the system. The overall quality and performance of the system was very good in terms of functionality, usability, and reliability.

*Recommendations* – It is recommended that future development such as the smartphone and tablet-based attendance monitoring should be integrated, a kiosk for grades and schedule viewing should also be placed inside the campus that is connected to the database server. Online student information systems should also be developed for the benefit of the students and parents, in easily monitoring school-related activities and requirements.

*Research Implications* – The study enabled the centralization of school and student data in storing, processing and retrieval. The System has been implemented in the college and has been updated now and then for continuous quality improvement.

# INTRODUCTION

Higher Education Institution (HEI's) implements Enterprise System (ES) as a backbone of its daily delivery of services (Rabaa'i et al., 2010), this ES is comprised of a wide range of transactions to name a few from Admission, Enrollment, Assessment, Subject Control, and Grade Management. Colleges and Universities acknowledge the importance of a School Management Information System (SMIS) because complex processes cannot be handled by certain people manually but made possible through the aid of technology, automating the school management information helps on reducing possible workloads, enhancing time management, and generate timely and quality reports (Shah, 2014). School administrators use different ICT Tools and Systems to carry out their administrative and management duties correctly and productively (Omotayo and Chigbundu, 2017). Many choose to rely on technology for managing data and information because it makes work more reliable and fast.

A Community College situated in the Northern Part of Mindanao Philippines has been established in the year 2005, since then the school depends on its transactions on a standalone program developed for the basic services of the school. Because the school lacks a centralized data center and a well-established network infrastructure, the majority of the information systems and systems generated are housed on a single personal computer and shared via a simple network arrangement.

The most common problems encountered by the school at the end of each semester are the student enrollment process, which typically lasts 2–3 weeks because most of the admission and storage of student information is done manually. Retrieval and generation of data and information is also a concern, as it takes a longer period to consolidate information that is required to complete the course.

By 2017 the program was redesigned and replaced by the NBCC-School Management System (NBCC_SMS) which is more centralized through a network-based environment by adopting client-server architecture to all of its systems and sub-systems. This system aids in the improvement of the school's enrollment process and data and information management by integrating a more centralized data storage system and addressing complex transactions by providing specific system modules to address the needs of each office in terms of data storage and retrieval.

The system has been improved from time to time as needs and demands grow because of the rapid increase of the population of the college. The school management information system uses a Hybrid System that runs two (2) platforms in its operation, one is a java-based environment for transactional processes and the other is a local web-based system for real-time queries and information retrieval. The school management system has been utilized for a variety of reasons, including data management, and the reports it generates aid in school decision-making for current and future advancements and innovations.

The School Management System has not been hosted online since the school has not purchased web hosting and a domain name, both of which are necessary for hosting systems such as the School Management System and the creation of a school webpage. As a result, the system can only handle transactions on the school's campus; online admission and enrolling are currently unavailable. The system has been implemented in the college and has been updated now and then for continuous quality improvement.

## LITERATURE REVIEW

The use of technology to update school facilities has a significant impact on student accomplishment; therefore, the need to develop an integrated School management system based on a centralized database will improve the quality of school services. (Balcita and Palaoag, 2020). This means that one of the most important aspects of developing and implementing efficient information systems for schools is to start with the genuine needs of the school, which include classroom needs and building educators. (Breiter and Light, 2006) and student needs (Reddy and Rathna, 2020).

A study of (Durnali, 2013) when comparing the data collection, processing, storage, accuracy, and analysis and dissemination of student data before and after an e-School was implemented in their school, it has been shown through his studies that there are improvements in terms of data collection, processing, storage, accuracy, and analysis and dissemination of student data. If technology is employed for school management and development, it has an impact on how society reflects the socioeconomic, cultural, and technical change (Yıkıcı et al., 2019).

Student information system (SIS) is a software program to handle student data for educational institutions. It is often known as the Knowledge Management System for Students (SIMS) (Murungan, 2020). Information systems are created to provide schools with a solution and feedback to encourage the effectiveness of learning, teaching, and administration purposes; this will exploit the school that needs the system (Kurniawan & Andika, 2019). A website-based information system boosts data administration efficiency, especially when it comes to student information. In order to achieve the goal of having an effective school management system, the system should have elements that improve the quality of schoolwork (Dewantara, Piarsa, & Buana, 2019).

Research on creative information management in Taiwan illustrated the value of the school management method. Innovative information management makes students more inspired, enhances their learning effectively, and increases the sense of classes and schools being defined (Chen et al., 2014).

A study in Turkey shows how teachers and the principal believe that the e-school system is adequate in terms of administrative relations, student affairs, and student report card work time (Polat & Arabaci, 2013). A developed web-based information system for school management provides leverage for schools that need the application to facilitate learning, teaching, and administration quality and effectiveness. It is always necessary to have a modern school management system information system (Pavlović, Ranđić & Paunović, 2014).

It is necessary to keep school records and manage them properly because it is an important aspect of the school's leverage in terms of keeping things in their rightful place to ensure quality processing and record-keeping. This will also help school managers in their decision-making process and to also enhance the implementation of usable records in schools that will lead to cost savings, transparency, easy accessibility, accountability, and retrieval of required information from their storage (Akinloye, Adu, & Ojo, 2017). The key areas that need to be present in a school information management system are open standards, interoperability, transition, accessibility, cost efficiency, statutory-based innovations, and usability (Strickley, 2011).

## METHODOLOGY

*Software Development*

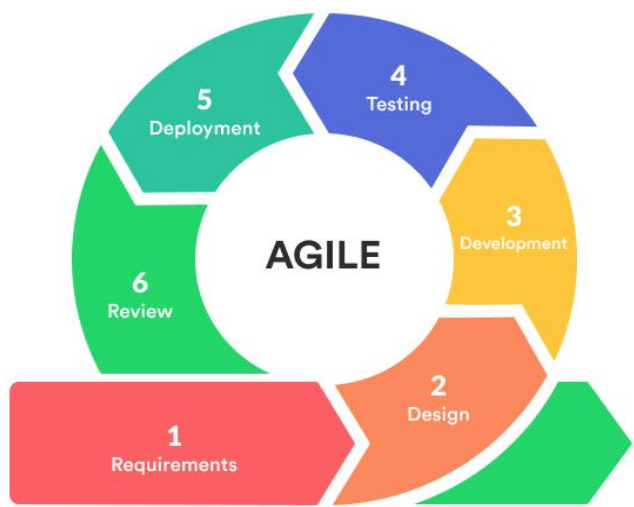

*Figure 1.* Agile Methodology in System Development
*source: Okeke(2021), retrieved from https://targettrend.com/agile-methodology-meaning-advantages-disadvantages-more*

Over time, the SDLC is the most reliable method of software development (Khan, Shadab, & Khan, 2020). However, new Agile approaches are gradually replacing traditional methods (Domingues, 2018). The Agile technique seeks to provide more value to consumers by reducing development periods and incorporating frequent modifications (Okeke, 2021). The Agile technique, as indicated in Figure 1, is ideally suited for speedy and effective software development due to its adaptive nature, early delivery, and flexible life cycle (Srivastava, Bhardwaj, & Saraswat, 2017).

*Requirements*

Schools processes and procedures require a lot of data to consider. There are a lot of methods in data collection and requirement gathering, the proponents decided to use two (2) methods, the primary method to use in this study are interviews with the key informant or the person who has the prime knowledge regarding the office transaction because it allows the proponents to ask open-ended questions which is beneficial on designing the User Interface (UI) and the Data Processes (Ainsworth, 2020). The second method is forms and reports collection, in this method Data documents were gathered such as registration forms, curriculum, sample syllabus, assessment form, grade slips, enrollment forms, and school reports.

*Design*

Those forms and reports gathered are analyzed in coming up with a functional design of the Network Infrastructure, a framework for the specification of a network's physical components and their functional organization and configuration, the System Architecture which is the conceptual model that defines the structure, behavior, and more views of a system, the Database Model that determines the logical structure of a database and fundamentally determines in which manner data can be stored, organized and manipulated, the User Roles that defines permissions for users to perform a group of tasks and the design of the User Interface (UI) used to build interfaces in software or computerized devices, focusing on looks or style.

*Development*

The system is a hybrid type of development because it consists of two development environments. One environment is Java and the other one is a web-based platform. The java platform acts as the transactional program which allows users to perform real-time school-based transactions such as Subject control, Enrollment, Assessment, and printing of reports, while the Web-Based System allows different user roles to interact with the system through admission, Student Information, summative reports such as class list, student grade, grade evaluation, and grade input for the faculty.

The system authenticates the user and identifies its privileges. There are three types of users namely, the Super Admin (System Administrator), the Staff (School Registrar, Admin Assistant, Program Heads), and the Instructors. The super admin supervises the overall functions of the system from managing default system settings to all records of the student. The super admin can enroll students, add entities such as rooms, instructors, subjects, sections, and user profiles, control assessment, generate reports, modify entity details such as school default settings.

The school registrar can only generate student grades reports and TOR can change student ID manually if in case necessary, can drop students, and can modify staff-privilege settings. The admin assistant produces the student assessment and COR and applies for scholarship programs if available. The admin assistant can also generate admin assistant reports such as grade slip, subject-enrolled by the student's summary, list of the payee and non-payee students, student assessment summary report. The program head can add subject info, section details, and instructors, can enroll students, and lastly, can generate student lists per course, department, and year level. The instructor can only upload their grades to the system through a manual procedure or file upload via (.xls) format provided by the system.

*Testing and Evaluation*

The respondents of the study were the faculty and staff of the community college in Northern Mindanao. They were chosen as respondents because of their involvement in the process of school information management. There were (4) registrar and admin staff, and (5) instructors for a total of (9) respondents. All of which are direct users of the school management information system.

The researchers utilized the purposive sampling technique. It was used to select the subjects that best fit the study by own judgment and used when dealing with a limited target of subjects. The testing was conducted on each respondent individually during their most convenient time. The researchers used ISO/IEC 25010:2011 in terms of functionality suitability, usability, and reliability. Table 1 shows the components of the evaluation questionnaire.

Table 1. Components of the evaluation questionnaire

| Criteria | Indicators |
|---|---|
| 1. Functionality Suitability | 1.1 Functional Completeness |
|  | 1.2 Functional Correctness |
|  | 1.3 Functional Appropriateness |
| 2. Usability | 2.1 Appropriateness Recognizability |
|  | 2.2 Learnability |
|  | 2.3 Operability |
|  | 2.4 User Error Protection |
|  | 2.5 User Interface Aesthetics |
|  | 2.6 Accessibility |
| 3. Reliability | 3.1 Maturity |
|  | 3.2 Availability |
|  | 3.3 Fault Tolerance |
|  | 3.4 Accountability |

To measure the quality of the system a 5-point scale was used. The average weighted mean was computed and interpreted as (1) 4.20-5.00: Excellent; (2) 3.40-4.19:

Very Good; (3) 2.60-3.39: Good; (4)1.80-2.59: Fair; and (5) 1.00-1.79: Poor. The results are presented in the succeeding sections of this paper.

*Implementation*

The system is hosted in a local server in the administration building which includes hosting the Database Management System, the Web Portal, and the Java-Based Platform. The server is connected across the College that connects one building to another through basic network Infrastructure using a Star Topology.

*Review*

Upgrades and updates are performed remotely in the IT department to ensure that concerns and problems that arise upon the continual use of the system are addressed right away. Data backup and restore are already included to avoid data loss if there is a failure of the hardware device.

## RESULTS AND DISCUSSION

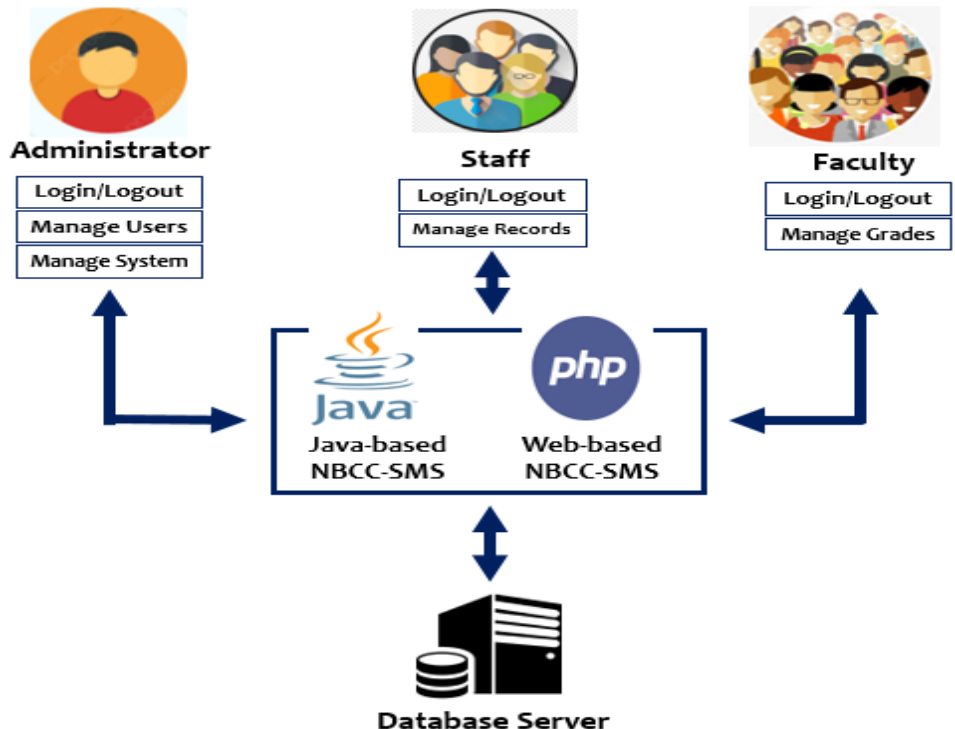

*Figure 2.* System Architecture of the e-school system

The system's users as shown in Figure 2 are the administrator, school staff, and faculty. Each has different intended modules. The administrator oversees the entire system and manages it. The administrator is the only authorized person to manage users in terms of adding new systems users for both staff and faculty.

The staffs include the registrar, guidance, library, and student affairs offices. The staff can only manage the records intended for their office. An audit trail is in place to keep track of the transactions made through the system. This will help the administrator and users in times of unprecedented scenarios. The faculty can manage the grades of their students such as viewing and uploading. All of the system transactions from each department run into the database server of the e-school system.

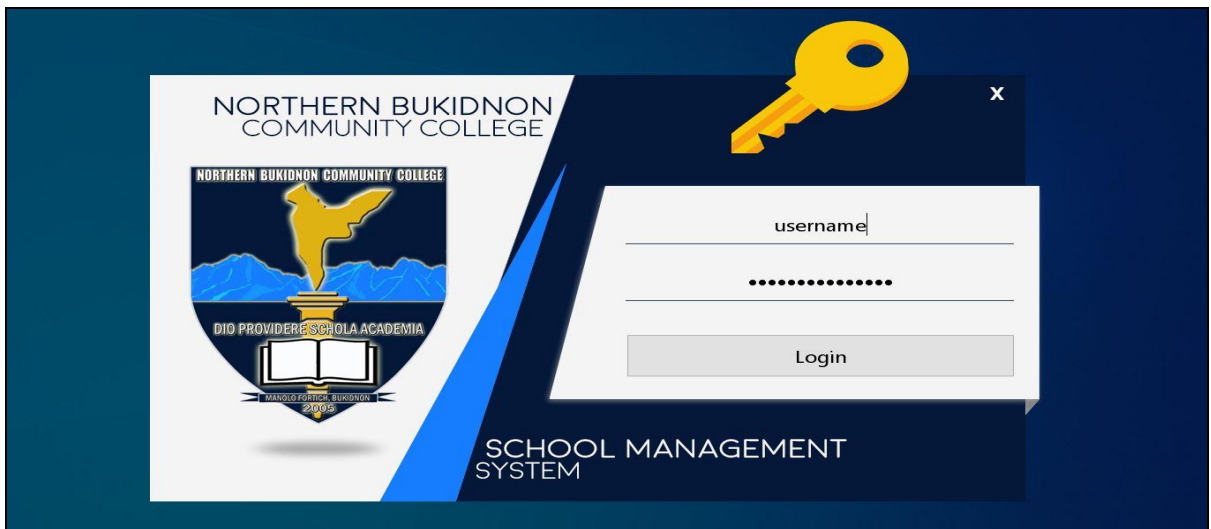

*Figure 3.* Login Module

The login module (Figure 3) is the main entrance to access the system. Only registered personnel can access the system to maintain exclusive accessibility. Three types of users can log in: The Administrator, College Personnel/Staff, and the Faculty. Accounts are registered by the administrator and then given to the authorized faculty and staff.

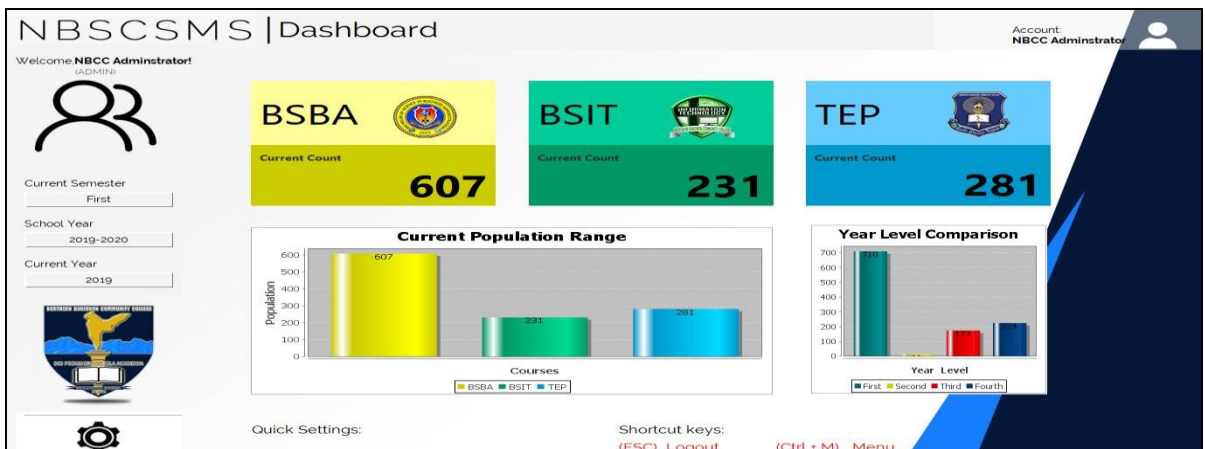

*Figure 4.* Dashboard

The NBCC-SMS Dashboard as shown in Figure 4 above summarizes all key functions necessary in one place. The left panel holds the information of the current semester as well as the current School Year to operate. A gear button is used to show other options such as adding new faculty, subjects, sections, and rooms, viewing of records, and enrollment of students. On the middle panel which the expanded panel shows the summary of enrollment disaggregated from the 3 different departments of the college to monitor the real-time updates of enrollment on a specific school year and semester.

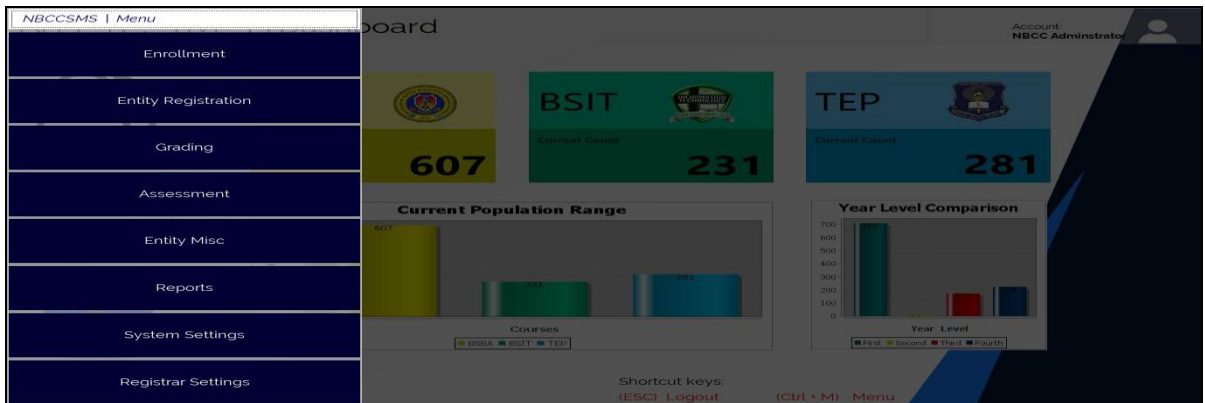

*Figure 5.* Dashboard and Menu

The Dashboard menu as shown in Figure 5, appears only when the gear button is clicked. This Menu contains the important sub-modules of the system which are fundamental requirements before student enrollment. The menu contains the following functional components: Enrollment, Entity Registration, Grading, Assessment, Entity Miscellaneous, Reports, System Settings, and Registrar's Settings.

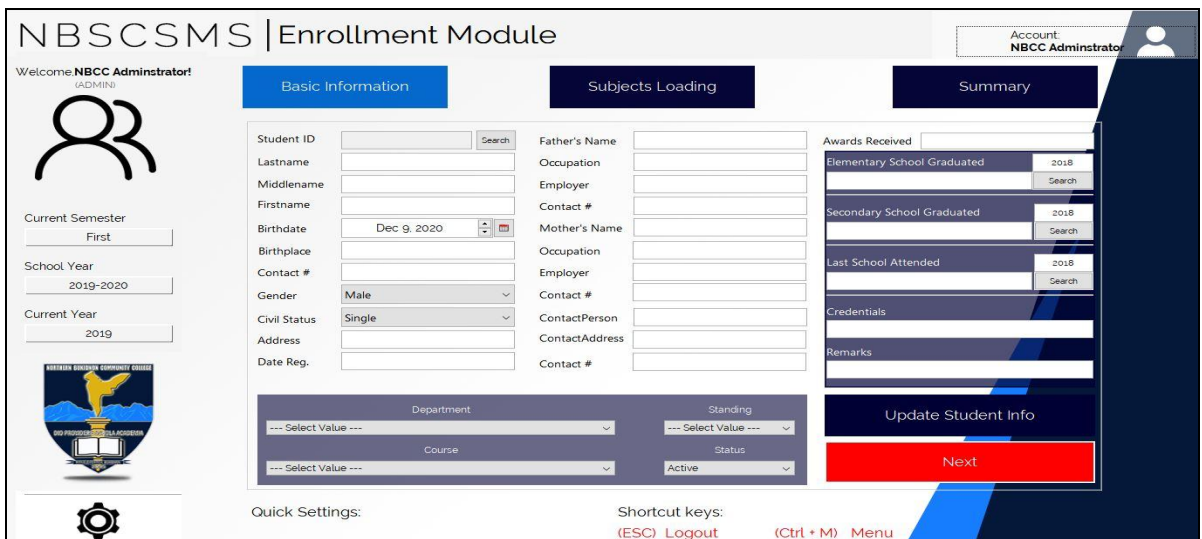

*Figure 6.* Enrollment Module

Figure 6 shows the Enrollment Module of the System. This module allows the Program Administrator to enroll students based on their evaluation for the previous semester. This module update student scholastic information such as Course and Year, Basic Information, and the previous school or course is taken or attended.

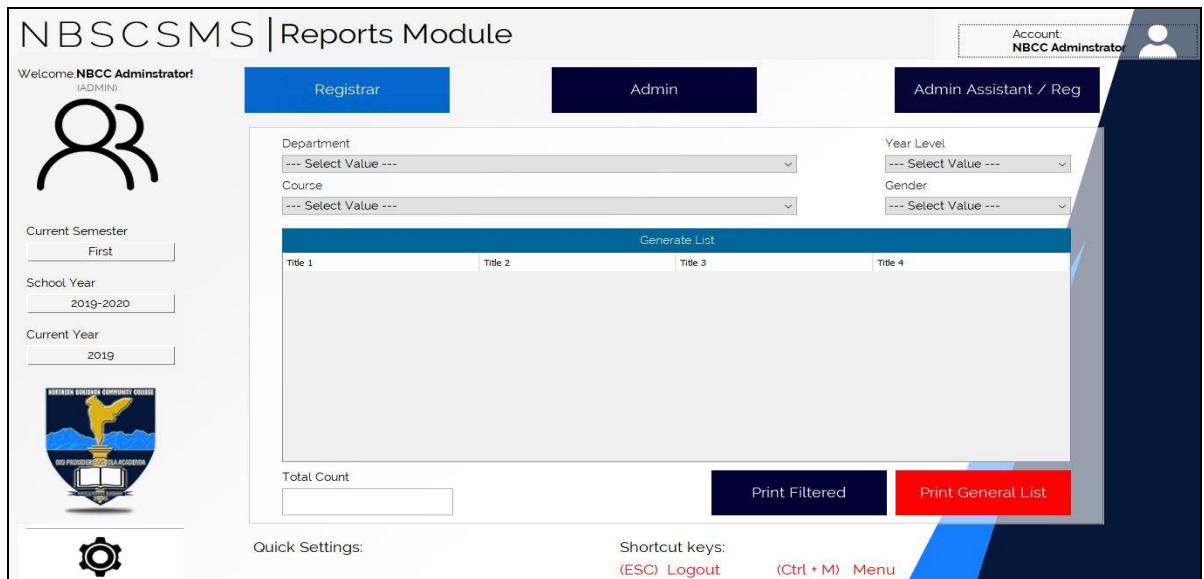

*Figure 7*. Reports Module

Figure 7 as shown consolidates the expected and necessary reports to be generated by the system. This includes the report needed by the registrar, the admin officer, and the admin assistant. One example to generate in this module is the information of how many students enrolled in a specific department, disaggregated values such as how many enrolled on a specific year level, gender, course, and major. The reports generated are considered an official basis for decision making and strategic plan of the school as well as report submissions to the Commission on Higher Education regional office.

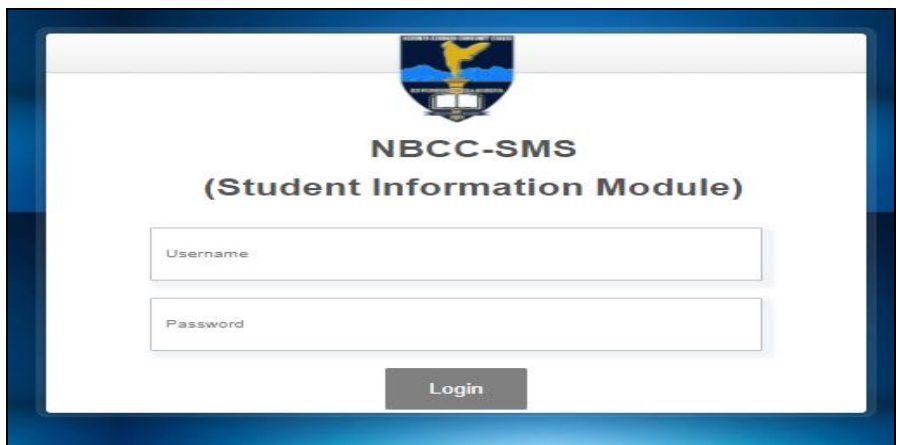

*Figure 8*. Login Module for the Web-based System

Figure 8 is another Login Module for the web portal of the School Management System. This web module allows department heads and administrators to see the overall student information and the school's real-time enrollment update. Only authorized personnel are allowed to access the web portal of the school because it contains confidential student records.

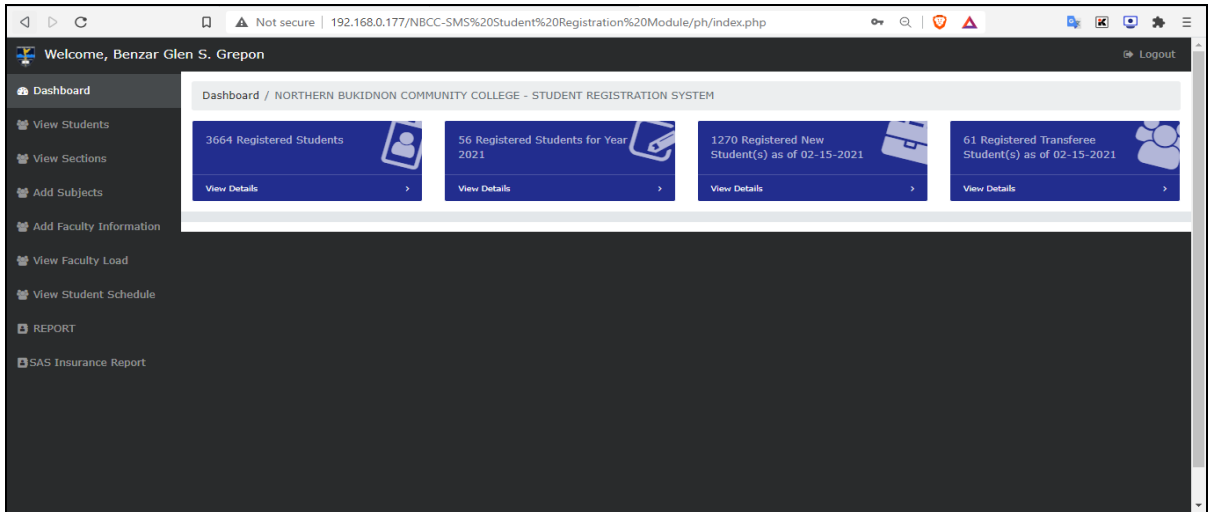

*Figure 9*. The Dashboard of the Web-based System

The Dashboard of the web portal in Figure 9 is similar to the dashboard in Figure 4 for the Java program since it holds the necessary functional attributes of the system. In the left portion of the window is the menu list that pertains to links to the specific page, the right panel holds the summary of information expected during the enrollment and the partial viewing of how many students enrolled, shifted, and transferred to the school.

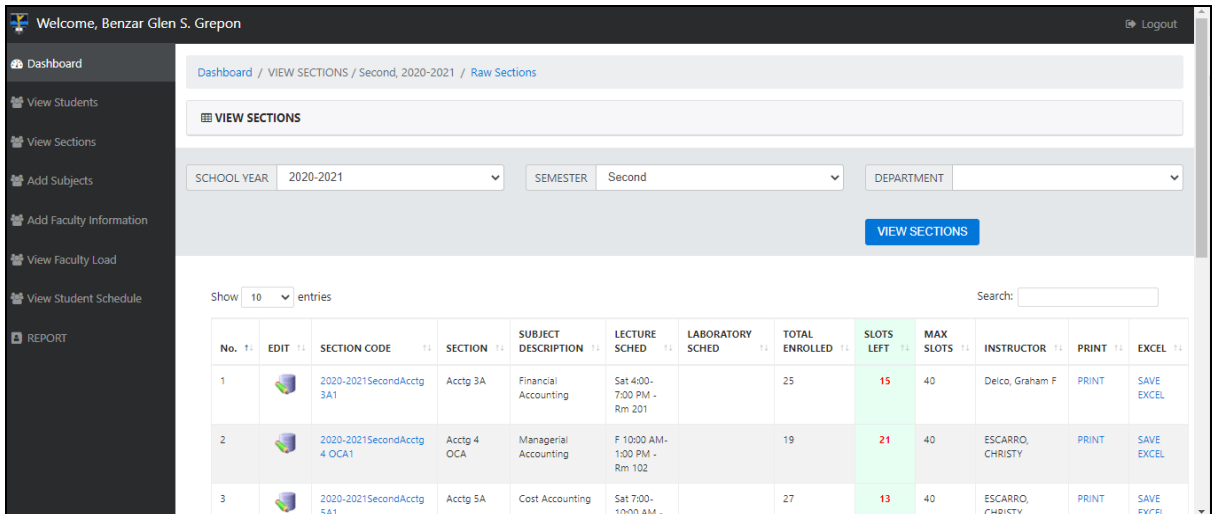

*Figure 10*. View Sections

Sections as shown in Figure 10 are the detailed information of a subject and its section needed to identify how many sections have students enrolled, how many remaining slots before it becomes full, the maximum slots and report generation, and download of an excel file summarizing the student list per sections.

*Figure 11.* Faculty Load

Faculty Load as shown in Figure 11 is the portal wherein a program coordinator will see the summarized subjects assigned to a faculty and know the details of the subject assigned to the faculty in monitoring the total number of units and the schedule to avoid conflicts. This portal was used by the department chairs to print the teacher's load report to be distributed to the faculty under his/her department.

*Figure 12.* Student Schedule

One important module in the system is the viewing of the student schedule as shown in Figure 12, after enrollment students would like to know the summary of their subjects. This module allows the department head to print the individual student load form of a student. The generated and printed form is used to track the academic status of the students as it is included in their 201 File.

*Figure 13.* Student Grades

Student grades shown in Figure 13 are one of the reports generated by this system. Authorized personnel should log in to the system because this type of information is readily accessed. During student evaluation before enrollment, department chairs review and print student grades to identify whether they are qualified to move on to the next year's level and get advanced subjects or they need to retake subjects that are not passed. Upon the request of the student to know their grades, department heads using this module can print a hard copy of this student's grade then distribute it to the student.

*Figure 14.* Sample Permanent Record of a Student

To know properly evaluate a student the summary of subjects taken as shown in Figure 14 is generated and printed by the department head. This report contains the overall academic subjects taken and enrolled by a student in the school. When a student would like to transfer to another school, this report is printed as a temporary grade evaluation while waiting for the official transcript of records issued by the Registrar's office.

*Evaluation results*

Table 2 shows the quality level of the e-schools management system. As observed from the table, the functionality suitability of the system has a 4.15 evaluated mean, which is interpreted as very good. This means that the system provides complete functionality that is accurate and appropriate.

Usability is graded with a mean of 4.15 which is interpreted as very good. This means that the system is easy to learn, use and navigate, and it has an intuitive design that also prevents error among users. Reliability has a mean of 3.83 which is interpreted as very good. This means that in terms of maturity, availability, fault tolerance, and recoverability the system ensures the information in the systems. The overall average weighted mean of the system evaluation in different aspects is 4.04 which is verbally interpreted as very good. This only proves the quality of the developed management information system for the school.

Table 2. Evaluation results summary.

| Criterion | Weighted Mean | Verbal Interpretation |
|---|---|---|
| Functionality Suitability | 4.15 | Very Good |
| Usability | 4.15 | Very Good |
| Reliability | 3.83 | Very Good |
| **Average Weighted Mean** | **4.04** | **Very Good** |

## CONCLUSIONS

The School Management System was designed and developed based on the needs of the school especially on its major transactions: Admission, Enrollment, Accounting, Student Information System, Grading, and Report Generation. Based on the preceding findings of the study, the respondents agreed that the developed e-school system was functional and lifted the transaction process of the school. The faculty and staff have benefited from making use of the system. The overall quality and performance of the system were very good in terms of functionality, usability, and reliability.

## RECOMMENDATIONS

It is recommended that future development such as a smartphone- and tablet-based attendance monitoring should be integrated, a kiosk for grades and schedule viewing should also be placed inside the campus that is connected to the database server. Online student information systems should also be developed for the benefit of the students and parents, in easily monitoring school-related activities and requirements.

# REFERENCES

Ainsworth, Q. (2020). *Data collection methods: jotform education.* Retrieved from https://www.jotform.com/data-collection-methods/

Akinloye, G. M., Adu, E. O., & Ojo, O. A. (2017). Record-keeping management practices and legal issues in the school system. *The Anthropologist, 28*(3), 197-207. doi: 10.1080/09720073.2017.1335832

Balcita, R. E., & Palaoag, T. D. (2020). Building a framework for the integration of school management systems (BFISMS). *International Journal of Information and Education Technology, 10*(6), 55-459. https://doi.org/10.18178/ijiet.2020.10.6.1406

Breiter, A., & Light, D. (2006). Data for school improvement: Factors for designing effective information systems to support decision-making in schools. *Journal of Educational Technology & Society, 9*(3), 206-217.

Chen, P. C., Lan, T. S., Lan, Y. H., & Hsu, H. Y. (2014). Dynamic effect of knowledge management system on school management. *Journal of Theoretical and Applied Information Technology, 61*(2), 249-253.

Dewantara, K. W., Piarsa, I. N., & Buana, P. W. (2019). Website-based high school management information system. *International Journal of Computer Applications Technology and Research, 8*(11), 420-424. doi: 10.7753/ijcatr0811.1003

Ribeiro, A., & Domingues, L. (2018). Acceptance of an agile methodology in the public sector. *Procedia computer science, 138*, 621-629.

Durnalı, M. (2013). The contributions of e-school, a student information management system, to the data processes, environment, education and economy of Turkey. *Proceedings of the Asian Conference on Technology in the Classroom,* (pp. 170-184). Retrieved from http://papers.iafor.org/wp-content/uploads/papers/actc2013/ACTC2013_0233.pdf

Kurniawan, Y., & Andika, A. (2019). Development of web based school management information system (a case study approach). *International Journal of Mechanical Engineering and Technology, 10*(2), 652-661.

Khan, M. E., Shadab, S. G. M., & Khan, F. (2020). Empirical study of software development life cycle and its various models. *International Journal of Software Engineering, 8*, 16-26.

Murugan, M. G. (2020). Student information system. *International Journal for Research in Applied Science and Engineering Technology, 8*(4), 9-12. https://doi.org/10.22214/ijraset.2020.4002

Okeke, N. (2021). *Agile Methodology: Meaning, advantages, disadvantages & more.* Retrieved from https://targettrend.com/agile-methodology-meaning-advantages-disadvantages-more/

Omotayo, F.O., & Chigbundu, M.C. (2017). Use of information and communication technologies for administration and management of schools in Nigeria. *Journal of*

*Systems and Information Technology, 19*(3/4), 183-201. https://doi.org/10.1108/JSIT-06-2017-0045

Pavlović, M. I. J., Ranđić, S., & Paunović, L. (2014). Information technologies in the contemporary school management system. *Emerald Emerging Markets Case Studies, 4*(6). https://doi.org/10.1108/EEMCS-11-2012-0195

Polat, M., & Arabaci, I. B. (2013). Evaluation of E-school applications as a management information system. *Elementary Education Online, 12*(2), 320–333. https://doi.org/10.17051/io.49201

Rabaa'i, A., Bandara, W., Gable, G. (2010). Enterprise Systems in Universities: A Teaching Case. American Conference on Information Systems. *In Proceedings of the 16th Americas Conference on Information Systems. Association for Information Systems (AIS).* 1-13.

Reddy, S., & Rathna, R. (2020). Android based Student Management System (No. 3018). EasyChair Preprint.

Shah, M. (2014). Impact of management information systems (MIS) on school administration: What the literatures says. *Procedia-Social and Behavioral Sciences, 16*(1), 2799-2804.

Srivastava, A., Bhardwaj, S., & Saraswat, S. (2017). SCRUM model for agile methodology. In P. N. Astya, A.Swaroop, V. Sharma, M. Singh, & K. Gupta (Eds.), *Proceedings of the International Conference on Computing, Communication, and Automation (ICCCA).* Paper presented at the 2017 International Conference on Computing, Communication, and Automation, India (pp. 864-869). India: IEEE.

Strickley, A. (2011). A baseline for a school management information system. In A. Tatnall, O. C. Kereteletswe, & A. Visscher (Eds.), *Information Technology and Managing Quality Education.* Paper presented at the IFIP Advances in Information and Communication Technology (ITEM 2010), vol. 348, Botswana (pp. 62-74). Berlin:Springer. doi: 10.1007/978-3-642-19715-4_7

Yıkıcı, B., Bastas, M., Altınay, F., Dagli, G., & Altınay, Z. (2019). The role of technology for school management and development. *Revista Inclusiones, 6,* 100-115.